\documentclass[fleqn,twoside]{article}
\usepackage{espcrc2}
\usepackage{dsfont}

\usepackage{graphicx}
\usepackage[figuresright]{rotating}

\newcommand{\be}{\begin{equation}}
\newcommand{\ee}{\end{equation}}
\newcommand{\ba}{\begin{array}}
\newcommand{\ea}{\end{array}}
\newcommand{\bea}{\begin{eqnarray}}
\newcommand{\eea}{\end{eqnarray}}

\newcommand{\tr}{^{\rm tr}}

\newcommand{\lt}{\left}
\newcommand{\rt}{\right}

\newcommand{\psibar}{{\overline\psi}}

\newcommand{\lesim}{${\lower 2pt\hbox{$\scriptstyle
<$}\atop\raise 4pt\hbox{$\scriptstyle\sim$}}$} 
\newcommand{\grsim}{${\lower2pt\hbox{$\scriptstyle >$} \atop\raise4pt\hbox 
{$\scriptstyle\sim$}}$}

\newcommand{\AmS}{{\protect\the\textfont2
  A\kern-.1667em\lower.8ex\hbox{M}\kern-.125emS}}

\title{The Lattice NJL Model at Non-zero Baryon and Isospin Densities}

\author{David N. Walters\address{Theoretical Physics Group, Department
        of Physics and Astronomy, University of Manchester,
        Manchester M13 9PL, U.K.}
        and Simon Hands\address{Department of Physics, University of Wales
        Swansea, Singleton Park, Swansea SA2 8PP, U.K.}\thanks{PPARC 
        Senior Research Fellow}} 

\begin{document}

\begin{abstract}
We present initial results of a numerical investigation of the
chiral symmetry restoring transition in the 
(3+1)-dimensional Nambu -- Jona-Lasinio model with both
non-zero baryon chemical potential ($\mu_B\neq0$) and isospin chemical
potential ($\mu_I\neq0$). 
With non-zero isospin chemical potential, the model
suffers from a sign problem. We proceed in two ways:
{\em (i)}
We perform ``partially quenched'' simulations in which $\mu_I$ is made
non-zero only during the measurement of chiral observables;
{\em (ii)}
We perform full simulations with imaginary isospin chemical potential
with the aim to analytically continue results to real $\mu_I$.
\vspace{1pc}
\end{abstract}

\maketitle

\section{Motivation}

At asymptotically high baryon chemical potential ($\mu_B$) and low
temperature ($T$) -- where QCD can be treated in a perturbative manner --
the ground-state of quark matter is found 
to be that of a colour-superconductor (for a recent review see
e.g.~\cite{Rischke:2003mt}). Unfortunately, the sign problem prevents
us from using lattice QCD to determine the ground-state at the more 
moderate densities typical in the cores of
compact (neutron) stars.

One way to proceed is to study model field theories such as the Nambu
-- Jona-Lasinio (NJL) model. This purely fermionic field theory, in
which colour-neutral quarks interact via a four-point contact term, not
only contains the same global symmetries as two flavour QCD, but can
be simulated on the lattice even with
$\mu_B\neq0$. In~\cite{Hands:2004uv} we show that the ground-state of
the lattice model with $\mu_u=\mu_d=\mu_B$, i.e. with ``up'' and ``down''
quarks sharing a common Fermi surface, exhibits s-wave superfluidity
via a standard BCS pairing between quarks of different flavours; i.e.
\be
\begin{array}{cc}
\lt<u d\rt>\neq0;&\Delta_{BCS}\neq0.
\end{array}
\ee

Within the cores of compact stars, however, the Fermi momenta $k_F^u$
and $k_F^d$ are expected to be separate. A simple argument
based on that of~\cite{Alford:1999pb} suggests that for a two flavour
Fermi liquid of massless quarks and electrons with $\mu_B=400{\rm MeV}$ and
both weak equilibrium ($\mu_d=\mu_u+\mu_e$) and charge neutrality
($2n_u/3-n_d/3-n_e=0$) enforced, all the Fermi momenta of the system 
are determined:
\bea
\ba{l}
k_F^u=\mu_u=\mu_B-\mu_e/2=355.5{\rm MeV},\\
k_F^d=\mu_d=\mu_B+\mu_e/2=444.5{\rm MeV},\\
k_F^e=\mu_e=89{\rm MeV}.
\label{eq:muIpred}
\ea
\eea

%
%


The effect of separating the Fermi surfaces of pairing quarks in QCD should
be to make the colour-superconducting phase less energetically
favourable. Introducing $\mu_I\propto(\mu_u-\mu_d)\neq0$
could prove a good method, therefore, to investigate
the stability of the superfluid phase.

\section{The Model}

The action of the lattice NJL model (with $a\to1$) is given by
\be\ba{r c l}
S_{NJL}&=&
\sum_{x y}\bar\Psi_x M[\Phi,\mu_B,\mu_I]_{x y}\Psi_y\\
&+&\frac2{g^2}\sum_{\tilde x}{\rm Tr} \Phi^\dagger_{\tilde x}\Phi_{\tilde x},
\ea\ee
where $\Psi\equiv(u,d)\tr$ is the $SU(2)$ doublet of staggered up and down
quarks defined on lattice sites $x$ and
$\Phi\equiv\sigma+i\vec\pi\do.\vec\tau$ is a matrix of bosonic
auxiliary fields defined on dual sites $\tilde x$.
The fermion kinetic matrix $M_{x y}$ is defined
in~\cite{Hands:2004uv} and we choose the same bare parameters used
therein. 

One can separate the Fermi surface of up and down quarks by
simultaneously setting baryon chemical potential
$\mu_B\equiv(\mu_u+\mu_d)/2\neq0$ and isospin chemical potential
$\mu_I\equiv(\mu_u-\mu_d)/2\neq0$. With $\mu_I=0$,
$\tau_2M\tau_2=M^*$, which is a sufficient
condition to show that $\det M$ is both real and
positive~\cite{Hands:2000ei}. With $\mu_I\neq0$ however, this is no
longer true such that once again we are faced with the sign problem.  

The fact that physically the two scales are ordered $\mu_I\ll\mu_B$
suggests that one may be able to apply techniques recently
developed to study QCD with $\mu_B\ll
T$~\cite{Fodor:2001,Allton:2002zi,deForcrand;DElia:2002}. First,
however, we present the results of a partially quenched calculation. 

\section{Partially Quenched $\mu_I$}

Whilst the primary motivation for investigating $\mu_I\neq0$ is to 
study the superfluid phase which sets in at large $\mu_B$, this requires one to
introduce an explicit symmetry breaking parameter ($j$); it is currently not
clear how to study $\mu_I\neq0$  in the $j\to0$ limit ~\cite{Hands:2004uv}.
Instead, we choose to study the chiral symmetry restoring phase
transition with the aim of controlling the systematics of introducing
$\mu_I\neq0$.  

\begin{figure}
\begin{center}
\includegraphics[width=6.6cm]{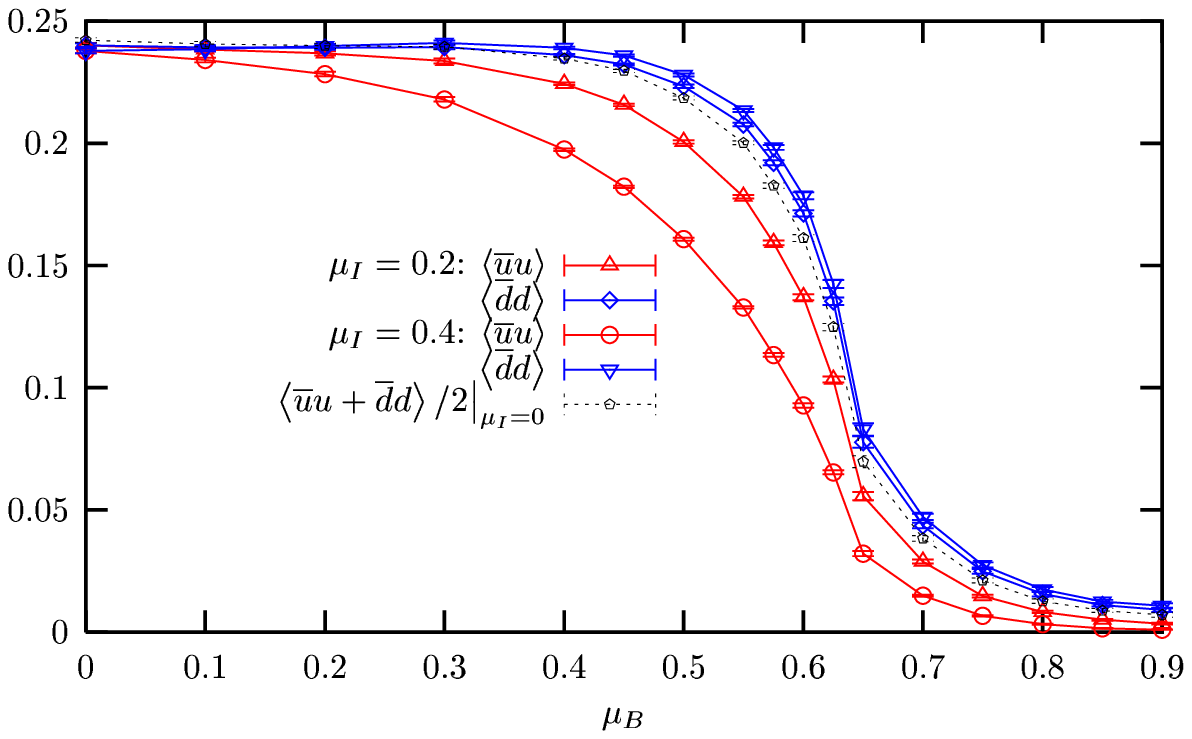}
\vspace*{-8ex}
\end{center}
\caption{$\left<\overline u u\right>$ and $\left<\overline d d\right>$
condensates for various $\mu_B$ and $\mu_I$ on a $12^4$ lattice.}
\label{fig:quench}
\vspace*{-4ex}
\end{figure}

The first step we take is to perform a ``partially quenched'' calculation
in which $\mu_I=0$ when generating the background fields and is made
non-zero only during the measurement of fermion
observables. In particular, we measure the up and down quark condensates 
\bea
\lt<\bar u u \rt>,\lt<\bar d d\rt>\equiv\frac1V\frac{\partial\ln{\cal
Z}}{\partial m_{u,d}}=
\frac{1}{2}\lt<{\rm t r}(\mathds{1}\pm\tau_3)M^{-1}\rt>
\eea
as functions of $\mu_B$ for various $\mu_I$ on a $12^4$ lattice. Some
results are presented in Fig.~\ref{fig:quench}. 

The results agree qualitatively with those of mean-field
studies of the model in which the introduction of a small
but non-zero $\mu_I$ is seen to suppress the up quark condensate and
enhance the down~\cite{Toublan:2003tt,Barducci:2004tt}. This can be
understood by noting that with $\mu_I\neq0$, $\mu_u$ ($\mu_d$) reaches the
critical chemical potential at a lower (higher) value of $\mu_B$. Why
$\lt<\bar u u\rt>$ deviates from the $\mu_I=0$ result more than
$\lt<\bar d d\rt>$ is not so clear, but may be an effect associated
with $T>0$ on a finite lattice~\cite{Barducci:2004tt}.

\section{Continuation from Imaginary $\mu_I$}

One recent method used to study QCD with small baryon chemical
potential $\mu_B$ is to measure quantities at imaginary $\mu_B$, where
the measure of the path integral is real, and fit results to a
truncated Taylor expansion in $\mu_B/T$ about $\mu_B=0$. One can then
analytically continue the results to real $\mu_B$~\cite{deForcrand;DElia:2002}. We propose to use a similar method by simulating at
imaginary isospin chemical potential ($\tilde\mu_I$), 
where once again $\det M$ is
real, and expanding observables in powers of $\tilde\mu_I/\mu_B$. 

For our initial investigation we have measured the quark condensates
and baryon and isospin number densities 
\be
n_{B,I}\equiv\frac1{2V}\frac{\partial\ln{\cal Z}}{\partial\mu_{B,I}}
=\frac14\lt<\bar u\gamma_0u\pm\bar d\gamma_0d \rt>
\ee
on a $12^4$ lattice at
$\mu_B=0.6$, which from Fig.~\ref{fig:quench} can be seen to be where
the effect of having $\mu_I\neq0$ is largest.
In QCD, one can show that e.g. $\lt<\psibar\psi\rt>$ expanded about
$\mu_B=0$ is analytic in $\mu_B^2$, such that for 
small imaginary $\mu_B$ the quantity remains real~\cite{Allton:2002zi}.
For our simulations, however, this is not the case, and measured
\begin{figure}
\begin{center}
\includegraphics[width=6.6cm]{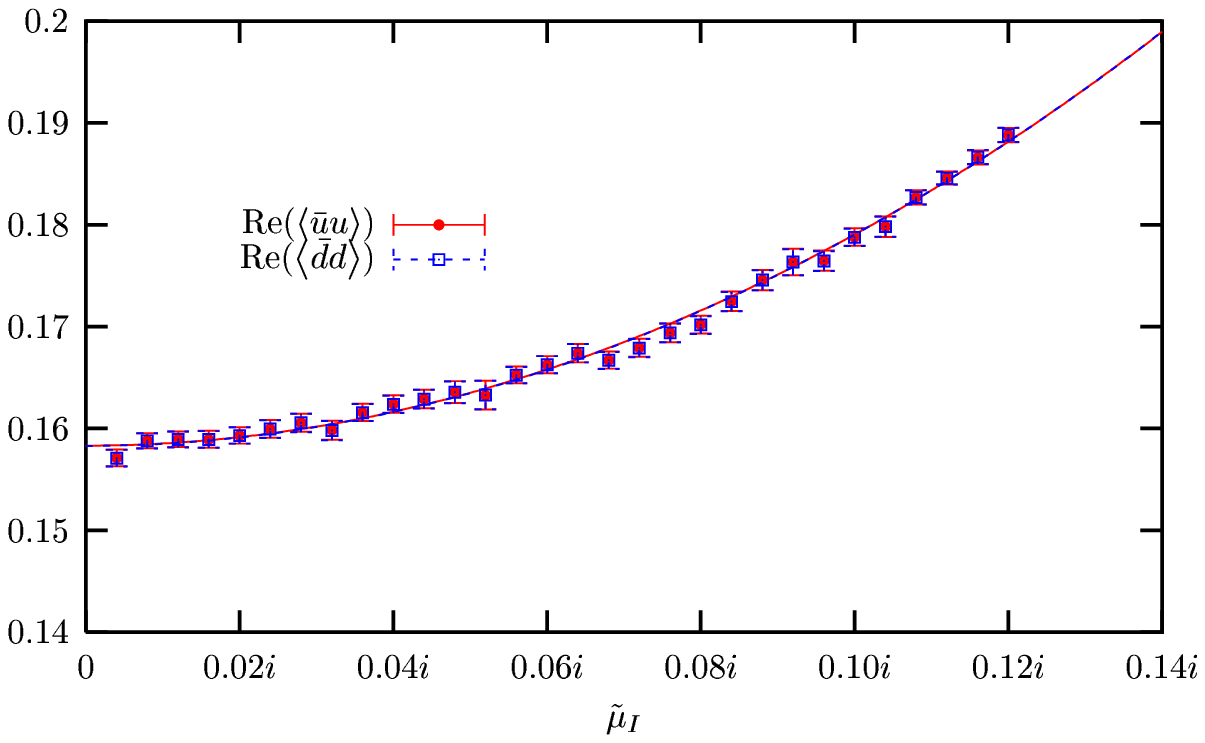}
\includegraphics[width=6.6cm]{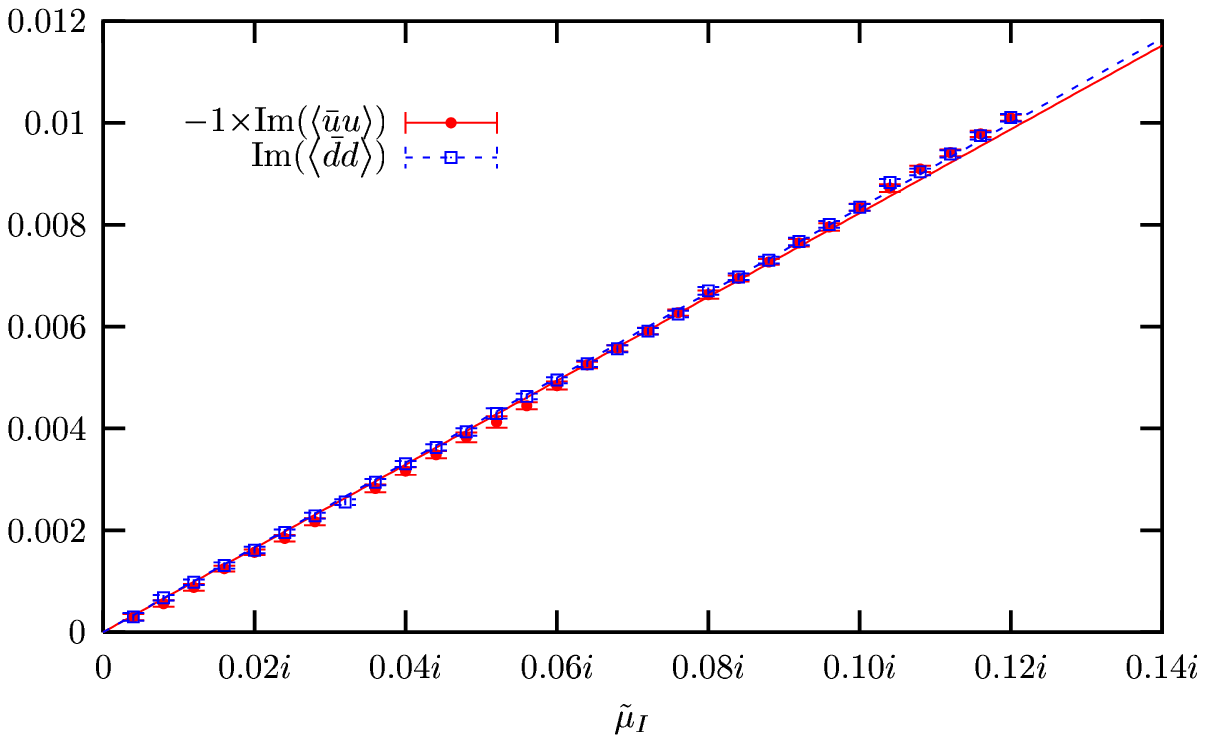}
\vspace*{-8ex}
\end{center}
\caption{Real and imaginary parts of $\left<\overline u u\right>$ and
$\left<\overline d d\right>$ as functions of imaginary $\mu_I$ with
$\mu_B=0.6$ on a $12^4$ lattice.}
\label{fig:pbp}
\vspace*{-4ex}
\end{figure}
quantities are, in general, complex. Therefore, we fit the data by
the Taylor series  
\be\lt(\ba{c}
\lt<\bar u u \rt>\\\lt<\bar d d \rt>\ea\rt)=
\sum_{n=0}^\infty\lt(\ba{c}A_n\\B_n\ea\rt)\lt(\frac{\tilde\mu_I}{\mu_B}\rt)^n
\label{eq:pbpexp}
\vspace*{-4ex}
\ee
and 
\be\lt(\ba{c}n_B\\n_I\ea\rt)=
\sum_{n=0}^\infty\lt(\ba{c}C_n\\D_n\ea\rt)\lt(\frac{\tilde\mu_I}{\mu_B}\rt)^n,
\label{eq:nexp}
\ee
each truncated at some suitable point. We can then analytically continue
to real $\mu_I$ using e.g.
\be
\lt<\bar u u\rt>=A_0+A_1\mu_I-A_2\mu_I^2-A_3\mu_I^3+\cdots
\ee

\begin{figure}
\begin{center}
\includegraphics[width=6.6cm]{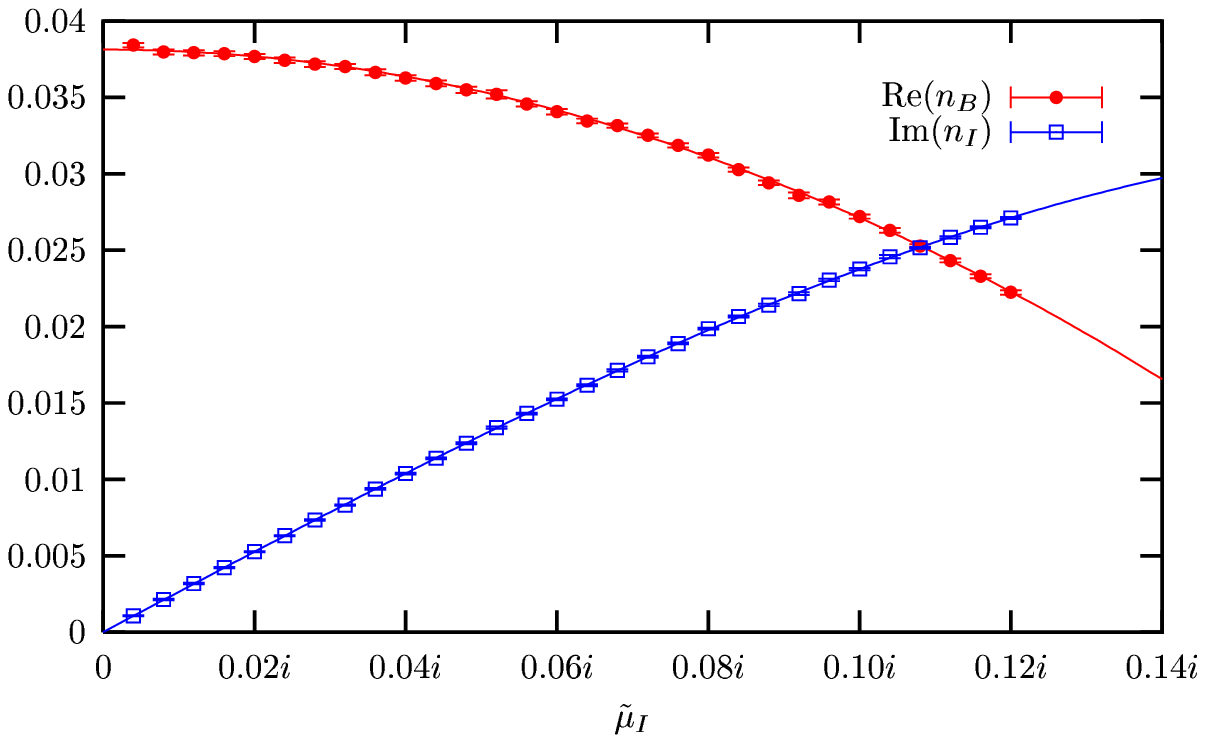}
\vspace*{-8ex}
\end{center}
\caption{Baryon and isospin density as functions of imaginary $\mu_I$ with
$\mu_B=0.6$ on a $12^4$ lattice.}
\label{fig:n}
\end{figure}

Figure~\ref{fig:pbp} shows the
real and imaginary parts of the condensates as functions of
$\tilde\mu_I$ with fits to {\em constant $+$ quadratic} and {\em
linear only}  forms
respectively. The quality of the fits is very good, suggesting 
that our ans\"atze are correct. The real parts are in complete agreement
whilst the imaginary parts are anti-correlated, which implies that the
sum of the condensates $\lt<\bar\psi\psi\rt>$ contains only the even
terms of the expansion, whilst their difference $\lt<\bar\psi\tau_3\psi\rt>$
contains only the odd terms.
A similar effect is seen in the number densities, as $n_B$ is found to
be real and therefore even in $\tilde\mu_I/\mu_B$, whilst the
isospin density is pure imaginary. For a good quality fit to
$n_I$, however, we have to include the cubic term. 
These data and fits are plotted in Fig.~\ref{fig:n}.

Whilst these results are only preliminary, we have shown that
 we can calculate the coefficients in (\ref{eq:pbpexp}) and
(\ref{eq:nexp}) as functions of $\mu_B$ and in
principle, reproduce reliable forms of the
curves in Fig.~\ref{fig:quench}. With this aim, we plan to repeat this
exercise for various values of $\mu_B$ in both the the chirally broken
and restored phases on various lattice volumes. Also, whilst
it is difficult to study the diquark 
condensate in the $j\to0$ limit, it would be interesting to compare
the response of $\lt< u d\rt>$ to $\mu_I$ at fixed $j$ to that of
$\lt<\psibar\psi\rt>$ at fixed mass.

\bibliographystyle{h-physrev3}
\bibliography{fermilab}
\end{document}